\documentstyle[12pt]{article}

\newcommand{\rA}{\rightarrow}
\newcommand{\ti}{\rightarrow \infty}
\newcommand{\tz}{\rightarrow 0}

\newcommand{\eb}{\makebox}

\begin{document}

\title{On the limits of Brans-Dicke spacetimes: a coordinate-free approach}
\author{
F. M. Paiva\thanks{
{\sc internet: fmpaiva@cbpfsu1.cat.cbpf.br;\ \ bitnet: fmpaiva@brlncc}}
\ \  and\ \
C. Romero\thanks{{\sc bitnet: cendif48@brufpb}} \\
\\
$^{\ast}~$Centro Brasileiro de Pesquisas F\'\i sicas, \\
R. Dr. Xavier Sigaud 150, \\
22290-180 Rio de Janeiro -- RJ, Brazil \\
\\
$^{\dagger}~$CP 5008 Departamento de F\'{\i}sica \\
Universidade Federal da Para\'{\i}ba, \\
58059-900 Jo\~ao Pessoa -- PB, Brazil}
\date{}
\maketitle
\begin{abstract}
We investigate the limit of Brans-Dicke spacetimes as $\omega\ti$
applying a coordinate-free technique. We obtain the limits of some
known exact solutions. It is shown that these limits may not correspond
to similar solutions in the general relativity theory.
\end{abstract}
{\sc pacs} numbers: 04.20.Jb, 04.20.Cv, 04.50.+h

\section{Introduction} \label{int} \setcounter{equation}{0}

In a recent paper \cite{RomeroBarros1993a} it has been showed that as
the scalar field coupling constant $\omega\ti$ some solutions of the
Brans-Dicke equations for a specified matter configuration (described
by the energy-momentum tensor $T_{\alpha\beta}$) do not go over to a
general relativity solution for the {\em same} $T_{\alpha\beta}$. This
fact raises some discussion about the idea found in the literature
(see, for example, \cite{Weinberg}) that the Brans-Dicke theory should
reduce to general relativity as $\omega\ti$.

Limits of spacetimes as some parameter varies were studied by Geroch
\cite{Geroch}, who called attention to the fact that the limit may
depend on the coordinate system used to perform the calculations. In
particular he showed that as the mass tends to infinity the
Schwarzschild solution may have as its limits the Minkowski space or a
Kasner spacetime. More recently Paiva et al. \cite{PaivaRM}
investigated this subject using a coordinate-free method.  Although
this new approach was developed in the framework of general relativity
it can be also applied to investigate the limits of Brans-Dicke
spacetimes.

In this paper we consider three exact solutions of Brans-Dicke field
equations. Using the coordinate-free technique we find the limits of
these solutions as $\omega\ti$. It turns out that these limit
spacetimes may be regarded as solutions of the Einstein general
relativity field equations for some specific matter tensors. Then we
compare these matter tensors with those of the original Brans-Dicke
solutions.

In section \ref{bd} we describe briefly the coordinate-free approach to
the problem of determining limits of spacetimes in general relativity and
show how it can be used to study Brans-Dicke spacetimes.

In section \ref{tn} we obtain the limits of the Nariai
\cite{Nariai1965a,Nariai1965b,Lorenz} and the O'Hanlon-Tupper
\cite{OHanlonTupper} solutions. These are Friedmann-Robertson-Walker
(FRW) cosmological models with euclidian spatial section in the
Brans-Dicke theory of gravity.

In section \ref{rb} we find the limits of a static FRW model
with closed spatial section, which is a vacuum solution of
Brans-Dicke equations with cosmological constant, found by
Romero and Barros \cite{RomeroBarros1993b}.

Finally, in the last section, we analyse the matter contents of
each limit spacetime obtained as solution of the general
relativity theory, drawing conclusions about the limits of
Brans-Dicke spacetimes as $\omega\ti$.

\section{The coordinate-free approach in Brans-Dicke's theory}
\label{bd} \setcounter{equation}{0}

The Brans-Dicke \cite{BransDicke} theory of gravity
assumes, as general relativity does, a lorentzian riemannian
manifold structure for the spacetime. The metric is determined
through the Brans-Dicke equations
\begin{eqnarray}
R_{\alpha\beta} - \frac{1}{2}g_{\alpha\beta}R  & = &
\frac{8\pi}{\phi}T_{\alpha\beta} + \lambda g_{\alpha\beta} + \label{bdeq} \\
&&\mbox{}+\frac{\omega}{\phi^2}\left(\phi_{,\alpha}\phi_{,\beta} -
\frac{1}{2}g_{\alpha\beta}\phi_{,\sigma}\phi^{,\sigma}\right) +
\frac{1}{\phi}\left(\phi_{,\alpha;\beta} - g_{\alpha\beta}\Box\phi\right)
\nonumber
\end{eqnarray}
and
\begin{equation}
\Box\phi
 \stackrel{\rm def}{=} \phi_{;\alpha}^{\ \ \alpha}
 = \frac{(8\pi T + 2\lambda\phi)}{2\omega+3}, \label{Box}
\end{equation}
where $\lambda $ is the cosmological constant, $\phi$ is the scalar field
(which plays the role of the inverse of a varying gravitational ``constant''),
$\omega$ is a parameter to be determined ``a posteriori'', $T_{\alpha\beta}$
is the matter tensor and $T$ is its trace. To preserve the equivalence
principle it is assumed that $\phi$ has no direct influence on the motion
of test particles, which does take place along geodesic lines. Formally, eqs.
(\ref{bdeq}) may be regarded as identical to the field equations of general
relativity if we define an ``effective'' energy-momentum tensor
$\tau_{\alpha\beta}$ by
\begin{equation}
\frac{8\pi}{G}\tau_{\alpha\beta} =
\frac{8\pi}{\phi}T_{\alpha\beta} +
\frac{\omega}{\phi^2}\left(\phi_{,\alpha}\phi_{,\beta} -
\frac{1}{2}g_{\alpha\beta}\phi_{,\sigma}\phi^{,\sigma}\right) +
\frac{1}{\phi}\left(\phi_{,\alpha;\beta} - g_{\alpha\beta}\Box\phi\right)
\end{equation}
subjected to equation (\ref{Box}). Therefore the coordinate-free
formalism used in \cite{PaivaRM} can be used here. In this
section we briefly review the main features of this formalism.

With the aim of solving the equivalence problem of metrics, Cartan
\cite{Cartan} showed that a spacetime is locally characterized by the
components of the Riemann tensor and its covariant derivatives
expressed in a constant frame (the {\em Cartan scalars}), up to
(possibly) the $10^{\rm th}$ order of derivation. Karlhede
\cite{Karlhede} introduced an algorithm used in practice to calculate
the Cartan scalars, which is implemented in the computer algebra system
{\sc sheep} \cite{Frick}, and constitute a suite of computer algebra
programs called {\sc classi} \cite{Aman}. A database of spacetimes
together with their Cartan scalars has been produced using {\sc classi}
\cite{ISCA}. To describe this algorithm we introduce the following
notation:
\begin{description}
\item $R_{n}$: the set of components of the covariant derivatives of the
Riemann tensor from the $0^{\rm th}$ to the $n^{\rm th}$ order --- the {\em
Cartan scalars};
\item $C_{n}$: the algebraically independent Cartan scalars among $R_{n}$;
\item $G_{n}$: the isometry group of $R_{n}$;
\item $H_{n}$: the isotropy group of $R_{n}$;
\item $t_{n}$: the number of independent functions of the coordinates in
$R_{n}$;
\item $d_{n}$: the dimension of the orbit of $G_{n}$, given by $d_{n} =
4 - t_{n}$. Note that ${\rm dim}(G_{n}) = d_{n} + {\rm dim}(H_{n})$.
\end{description}

The $0^{\rm th}$ step of the algorithm is the following. Choose a tetrad basis
in which the metric is constant. Calculate $R_{0}$ and, using Lorentz
transformations, find a basis where it takes a well-defined standard form
\cite{Karlhede}. Then find $H_{0}$ and $d_{0}$. $H_{0}$ is found with the help
of the Petrov and Segre classifications, while $d_{0}$ is the dimension of the
spacetime minus $t_{0}$. The next $n$ steps are similar to each
other, as described below.

Calculate $R_{n}$ and, using $H_{n-1}$, find a basis where $R_{n}$ takes a
well-defined standard form \cite{Karlhede}. Then find $H_{n}$ and $d_{n}$.
Now, $H_{n}$ is easily found as it must be a subgroup of $H_{n-1}$, which is
usually of low dimension. If $H_{n} \subset H_{n-1}$ (strictly, i.e.
$H_{n} \neq H_{n-1}$) or $d_{n} < d_{n-1}$, the
algorithm continues for another step. If $H_{n} = H_{n-1}$ and $d_{n} =
d_{n-1}$, then $R_{n}$ can be expressed as functions of $R_{n-1}$, and so the
algorithm stops. In most cases the algorithm stops at the $2^{\rm nd}$ or
$3^{\rm rd}$ covariant derivative --- the $4^{\rm th}$ is the highest
derivative necessary until now \cite{Koutras}.

Not all components of the Riemann tensor and its covariant derivatives are
algebraically independent; working with spinors MacCallum and {\AA}man
\cite{AmanMac} found a minimal set of algebraically independent quantities
which are actually calculated in the algorithm.  Their components will be
called {\em algebraically independent Cartan scalars} and denoted by $C_n$,
according to the order of derivation. For conformally flat solutions, as is
the case of the present paper, $C_2$ is:
\begin{eqnarray}
0^{\rm th}\ {\rm derivative}:& \Phi_{AB'}, & A,B = (0,~1,~2);
\label{c0a} \\
& \Lambda; & \label{c0b} \\
1^{\rm st}\ {\rm derivative}:& \nabla\Phi_{AB'}, & A,B = (0,\ldots,3);
\label{c1a} \\
& \nabla\Lambda_{AB'}, & A,B = (0,~1); \label{c1b} \\
2^{\rm nd}\ {\rm derivative}:& \nabla^2\Phi_{AB'}, & A,B = (0,\ldots,4); \\
& \Box\Phi_{AB'}, & A,B = (0,~1,~2); \\
& \Box\Lambda &
\end{eqnarray}
where $\Phi_{AB'}$ is the Ricci spinor, $\Lambda$ is the curvature scalar
spinor, $\nabla\Phi_{AB'}$ and $\nabla^2\Phi_{AB'}$ are the first and second
symmetrized covariant derivatives of $\Phi_{AB'}$, $\nabla\Lambda_{AB'}$ is
the first symmetrized covariant derivative of $\Lambda$ and $\Box\Phi_{AB'}$
and $\Box\Lambda$ are the D'Alembertians of $\Phi_{AB'}$ and $\Lambda$
\cite{ISCA}.

In order to study the limits of a spacetime we first find the Cartan scalars.
In general they are functions of the coordinates and the parameters of the
metric. Eliminating the coordinates and parameters from as many Cartan scalars
as possible (see eqs. (\ref{e0a})--(\ref{e1f}) below) they can be grouped in
three sets. The Cartan scalars on the first set are functions of the
coordinates and of the parameters. Those on the second depend on the
parameters and on the elements of the first set. Finally, the Cartan scalars
on the third set are functions of the elements on the first and second ones.
The expressions in the third set are parameter-free and coordinate-free,
therefore they hold for any limit of the spacetime (in a given basis). The
expressions in the second are coordinate-free but not parameter-free,
therefore their limits depend on the limits of the Cartan scalars in the first
set and the limits we choose for the parameters. The limits of the Cartan
scalars on the first set can be made arbitrary with a suitable choice of
coordinate system. Briefly, to obtain the limits of a spacetime in a given
basis we choose a limit to the Cartan scalars in the first set and to the
parameters. Then we calculate the limits of the Cartan scalars in the second
and third sets. With the limit of all the Cartan scalars we have the complete
characterization of the limit spacetime.

Nevertheless, this does not mean that the limit Cartan scalars one gets
correspond automatically to any metric, in other words, there is no {\em
a priori} guarantee that the Cartan scalars we obtain correspond to a
Riemann geometry. It is necessary to check whether they satisfy
integrability conditions \cite{KarlhedeLindstron}. This is usually very
hard. Although there is no algorithm to prove that a given set of
Cartan scalars correspond to a Riemann geometry, in most practical
cases we can either prove they do not satisfy the integrability
conditions or we can find the corresponding metric in the {\sc classi}
database we mentioned before.

Through the steps of the Karlhede's algorithm either the dimension of the
orbit or the dimension of the isotropy group decreases, or both.  Whenever the
isotropy group decreases, some freedom on the choice of the basis is lost.
Frame transformations could lead to new limits in the same way coordinates
changes do. Therefore, fixing the frame we may loose some limits. To recover
these limits we apply the lost isotropy freedom to the Cartan scalars. On the
zero derivative step of the algorithm this is done with the help of the Petrov
and Segre classifications. On the other steps, although a classification
scheme does not exist for the derivatives, the isotropy freedom is usually of
low dimension \cite{PaivaRM}.

\section{Nariai and O'Hanlon-Tupper spacetimes} \label{tn}
\setcounter{equation}{0}

The Nariai \cite{Nariai1965a,Nariai1965b,Lorenz} and the O'Hanlon-Tupper
\cite{OHanlonTupper} line element (with $k=0$) and scalar field may be
collectively written as
\begin{eqnarray}
ds^{2} &=& dt^2 - A^2\left(dr^2 + r^2 d\theta^2 +
         r^2\sin^{2}\theta d\phi^2\right),  \label{ds2} \\
\phi & = & \phi_0t^P, \label{fi}
\end{eqnarray}
where
\begin{eqnarray}
A & = & A_0t^Q, \label{A} \\
  &   & \nonumber \\
A_0 & \mbox{and} & \phi_0 \mbox{\ \ are constants,} \\
  &   & \nonumber \\
Q & = &
        \left\{ \begin{array}{ll}
        (3\omega+4)^{-1}\left(\omega+1\pm\sqrt{\frac{2\omega+3}{3}}\,\right)
        & \mbox{for O'Hanlon-Tupper,} \\
        & \\
        \frac{2+2\omega(2 - \gamma)}{4+3\omega\gamma(2-\gamma)}
        & \mbox{for Nariai,}
        \end{array}\right. \label{Q} \\
  &   & \nonumber \\
P & = &
        \left\{ \begin{array}{ll}
        {(3\omega+4)^{-1}\left(1\mp\sqrt{3(2\omega+3)}\,\right)}
        & \mbox{for O'Hanlon-Tupper,} \\
        & \\
        {\frac{2(4-3\gamma)}{4+3\omega\gamma(2-\gamma)}}
        & \mbox{for Nariai.}
        \end{array}\right. \label{P}
\end{eqnarray}
Both are solutions of the Brans-Dicke field
equations (\ref{bdeq}) with a vanishing cosmological constant.
The Nariai metric is a perfect fluid solution with a equation of state given
by
\begin{equation}
p = (\gamma-1)\rho,\mbox{\ \ where\ \ } 1 \leq \gamma \leq 2. \label{eqstat}
\end{equation}
The O'Hanlon-Tupper metric is a vacuum solution. In the null tetrad
\begin{equation} \begin{array}{lcl}
\omega^{0} &=& 1/\sqrt{2} (dt+Adr) , \\
\omega^{1} &=& 1/\sqrt{2} (dt-Adr), \\
\omega^{2} &=& 1/\sqrt{2}Ar (d\theta+i\sin(h) d\phi) , \\
\omega^{3} &=& 1/\sqrt{2}Ar (d\theta-i\sin(h) d\phi) ,
\end{array} \label{tetrada}
\end{equation}
the nonzero algebraically independent Cartan scalars ($C_n$) may be written as:
\begin{eqnarray}
\eb[12em][r]{$0^{\rm th}$ derivative:\hfill
$\Phi_{00'}$} & = & 2\Phi_{11'}, \label{e0a} \\
\eb[12em][r]{$\Phi_{11'}$} & = & \frac{1}{4}Qt^{-2}, \label{e0b} \\
\eb[12em][r]{$\Phi_{22'}$} & = & 2\Phi_{11'}, \label{e0c} \\
\eb[12em][r]{$\Lambda$} & = & \frac{1}{4}Qt^{-2}(2Q-1) \nonumber \\
& = & \Phi_{11'}(2Q-1), \label{e0d} \\
& & \nonumber \\
\eb[12em][r]{$1^{\rm st}$ derivative:\hfill
$\nabla\Phi_{00'}$} & = & \nabla\Phi_{33'}, \label{e1a} \\
\eb[12em][r]{$\nabla\Phi_{11'}$} & = &
\frac{1}{3}\nabla\Phi_{33'}, \label{e1b} \\
\eb[12em][r]{$\nabla\Phi_{22'}$} & = &
\frac{1}{3}\nabla\Phi_{33'}, \label{e1c} \\
\eb[12em][r]{$\nabla\Phi_{33'}$} & = &
- \frac{1}{\sqrt{2}}Qt^{-3}(Q+1) \nonumber \\
& = & -\frac{8}{\sqrt{2}}\frac{Q+1}{\sqrt{Q}}(\phi_{11'})^{\frac{3}{2}}
\nonumber \\
& = & - \frac{4\Phi_{11'}(\Lambda+3\Phi_{11'})}{\sqrt{\Lambda + \Phi_{11'}}},
\label{e1d} \\
\eb[12em][r]{$\nabla\Lambda_{00'}$} & = &  \nabla\Lambda_{11'}, \label{e1e} \\
\eb[12em][r]{$\nabla\Lambda_{11'}$} & = &
- \frac{1}{2\sqrt{2}}Qt^{-3}(2Q-1) \nonumber \\
& = &  -\frac{4}{\sqrt{2}}\frac{2Q-1}{\sqrt{Q}}(\phi_{11'})^{\frac{3}{2}}
\nonumber \\
& = &  - \frac{4\Phi_{11'}\Lambda}{\sqrt{\Lambda+\Phi_{11'}}}. \label{e1f}
\end{eqnarray}
We wrote as many scalars as possible as function of only few scalars.
$\Phi_{11'}$ is the only scalar which is function of $t$ and $Q$.
$\Lambda$ is function of $Q$ and $\Phi_{11'}$ only. All the others are
functions of $\Phi_{11'}$ and $\Lambda$. Therefore, once limits for
$\Phi_{11'}$ and $Q$ are chosen, the limits of the other scalars have to
satisfy the coordinate-free expressions above.

Some useful information such as Petrov and Segre types and dimension of the
isometry and isotropy groups can be found directly from the Cartan scalars. At
the zero order derivative step it is found that the Petrov type is 0, as the
Weyl spinor vanishes, and that the Segre type is [(111),1]. The isotropy group
$H_{0}$ is the 3-dimensional group SO(3). There is only one independent
function of the coordinates ($\Phi_{11'}$), so $t_{0} = 1$ and $d_{0} = 3$. At
the first derivative step we find that the isotropy group $H_{1}$ is the same
as $H_0$ and that there is no new independent function of the coordinates.
Therefore the algorithm stops.

The possible limits for $\Phi_{11'}$ are (1) 0; (2) a function of the four
coordinates, $f(x^i)$; (3) a non-zero constant; or (4) $\infty$. This last
case is a singular limit and shall not be consider in this paper. We shall
analyse the first three possibilities together with the limits of $Q$ which
are obtained from (\ref{Q}), namely\footnote{Note that Nariai metric does not
depend on $\omega$ if $\gamma = 2$. Therefore, this particular case will not
be considered.}
\begin{equation} \label{Qlim}
\lim_{\omega\ti}Q \rA
   \left\{ \begin{array}{lll}
   \frac{1}{3}            &                & \mbox{for O'Hanlon-Tupper,} \\
   & \\
   \frac{2}{3\gamma} & \gamma \neq 2 & \mbox{for Nariai}.
\end{array}
\right.
\end{equation}
\begin{enumerate}
\item $\Phi_{11'}\tz$. All Cartan scalars tend to zero, therefore the limit is
Minkowski space. Regarded as a general relativity solution, this corresponds
to vacuum.
\item $\Phi_{11'}\rA f(x^i)$. From (\ref{e0a})--(\ref{e1f}) the other Cartan
scalars have finite limits. In principle this could lead to many different
limit spacetimes, depending on the function $f(x^i)$.  Nevertheless any such
functions are related to each other by coordinate transformations, therefore
the respective spacetimes are equivalent (see also \cite{PaivaRM}). These
Cartan scalars correspond to the FRW metric with euclidian spatial section.
The line element is given by eq. (\ref{ds2}), where
\begin{equation}
A = A_0t^{Q'},\ \ Q'= \mbox{const.}
\end{equation}
Regarded as a general relativity
solution, it can be obtained from the field equations with  vanishing
cosmological constant and a perfect fluid matter tensor satisfying a
equation of state $p=(\gamma'-1)\rho$ where $\gamma'$ and $Q'$ are
related by (see \cite{KSMH})
\begin{equation}
Q' = \frac{2}{3\gamma'}. \label{Qlim2}
\end{equation}
\item $\Phi_{11'}\rA const \neq 0$. From the set of Cartan scalars
obtained in this limit one concludes that the corresponding spacetime
is conformally flat and of Segre type [1,(111)], has a transitive
seven-dimension group of isometry, is homogeneous and its isotropy
group is the SO(3). It then follows (see \cite{KSMH}) that the metric
is a static Robertson-Walker, i.e., the Einstein metric with positive
or negative tri-curvature. Nevertheless, for both these metrics, the
first order Cartan scalars can be shown to be identically zero. As from
eq. (\ref{e1a})--(\ref{e1f}) the first order Cartan scalars of the limit
are not identically zero one concludes that this set of Cartan scalars
does not correspond to a riemannian geometry, i.e., it does not satisfy
some integrability condition. So $\Phi_{11'}\rA const \neq 0$ is not a
possible limit for the metric (\ref{ds2}).
\end{enumerate}

This exhausts the [(111),1] Segre type limits. We shall now study the
possibility of limits with a different Segre type. Similarly to the
Petrov type specializations studied in \cite{Geroch,PaivaRM}, the Segre
classification satisfies a specialization scheme
\cite{Penrose1972,Hall1985,SanchezPlebanskiPrzanowski}. According to
this scheme, the possible limits of the Segre type [1,(111)] are
[(112)] and [(1111)]. The last one corresponds to $\Phi_{AB'}$
identically zero. From (\ref{e0d}) this implies in $\Lambda = 0$ and
therefore to Minkowski space which was already covered by our previous
analysis. We proceed now with the analysis of the Segre type [(112)]
limits.

The standard frame we used here (see eqs. (\ref{e0a})--(\ref{e0c})) for the
Segre type [1,(111)] is given by the following condition:
\begin{equation}
\Phi_{00'} = \Phi_{22'} = 2\Phi_{11'} \label{s1111}
\end{equation}
where the other components of $\Phi_{AB'}$ vanish. The only limits that can be
obtained on this frame are of the same Segre type or of Segre type [(1111)].
In order to study Segre type [(112)] limits, the frame has to be changed. A
suitable standard frame for Segre type [(112)] is given by the condition
\cite{JolyMacCallum}
\begin{equation}
\Phi_{22'} = 1  \label{s112}
\end{equation}
where the other components of $\Phi_{AB'}$ vanish. To transform our
previous frame to one in which the Cartan scalars might satisfy
condition (\ref{s112}) on the limits we use a boost
\begin{equation} \label{boost} \left( \begin{array}{cc}
1/z & 0 \\
0 & z \\
\end{array} \right), \end{equation}
where $z$ is real. The transformed Cartan scalars may be written as
\begin{eqnarray}
\eb[12em][r]{$0^{\rm th}$ derivative:\hfill
$\tilde{\Phi}_{00'}$} & = & 2z^{-4}\Phi_{11'}, \label{t0a} \\
\eb[12em][r]{$\tilde{\Phi}_{11'}$} & = & \Phi_{11'}, \label{t0b} \\
\eb[12em][r]{$\tilde{\Phi}_{22'}$} & = & 2z^4\Phi_{11'}, \label{t0c} \\
\eb[12em][r]{$\tilde{\Lambda}$} & = & \Lambda, \label{t0d} \\
 & & \nonumber \\
\eb[12em][r]{$1^{\rm st}$ derivative:\hfill
$\widetilde{\nabla\Phi}_{00'}$} & = & z^{-6}\nabla\Phi_{33'}, \label{t1a} \\
\eb[12em][r]{$\widetilde{\nabla\Phi}_{11'}$}
& = & 1/3z^{-2}\nabla\Phi_{33'}, \label{t1b} \\
\eb[12em][r]{$\widetilde{\nabla\Phi}_{22'}$}
& = & 1/3z^{2}\nabla\Phi_{33'}, \label{t1c} \\
\eb[12em][r]{$\widetilde{\nabla\Phi}_{33'}$}
& = & z^{6}\nabla\Phi_{33'}, \label{t1d} \\
\eb[12em][r]{$\widetilde{\nabla\Lambda}_{00'}$}
& = & z^{-2}\nabla\Lambda_{11'}, \label{t1e} \\
\eb[12em][r]{$\widetilde{\nabla\Lambda}_{11'}$}
& = & z^{2}\nabla\Lambda_{11'}, \label{t1f}
\end{eqnarray}
where $\Phi_{11'}$, $\Lambda$, $\nabla\Phi_{33'}$ and $\nabla\Lambda_{11'}$
are given by (\ref{e0b}), (\ref{e0d}), (\ref{e1d}) and (\ref{e1f}).

In order to obtain the standard form (\ref{s112}), the limits of $\Phi_{11'}$
and $z$ must be
\begin{equation} \label{lia} \begin{array}{ccccccc}
\Phi_{11'} & \rA & 0 & ~~{\rm and}~~ & z & \rA & \infty,
\end{array} \end{equation}
in such a way that the limit of the product in (\ref{t0c}) is
\begin{equation} \label{lib} \begin{array}{ccc}
2z^4\Phi_{11'} & \rA & 1.
\end{array} \end{equation}
These conditions hold if and only if
\begin{equation}
z^4 \rA \frac{1}{2\Phi_{11'}}, \label{zlimit}
\end{equation}
which is a coordinate-free definition.

Using this result and the limits (\ref{Qlim}) in (\ref{t0d})--(\ref{t1f}) we
find that in the limit the nonzero Cartan scalars are
\begin{eqnarray}
\eb[12em][r]{$0^{\rm th}$ derivative:\hfill
$\tilde{\Phi}_{22'}$} & \rA & 1, \label{l0a} \\
 & & \nonumber \\
\eb[12em][r]{$1^{\rm st}$ derivative:\hfill
$\widetilde{\nabla\Phi}_{33'}$} & \rA & -2\frac{Q+1}{\sqrt{Q}} .
\label{l1a}
\end{eqnarray}
This set of Cartan scalars gives rise to Segre type [(112)] metrics with the
isotropy groups $H_0 = H_1$ equal to a 2-parameter group of null rotation plus
a 1-parameter group of spatial rotations. $d_0 = d_1 = 4$ as the Cartan
scalars are constant. Therefore the spacetime is ST-homogeneous.

Searching the {\sc classi} database \cite{ISCA} for spacetimes with those
Cartan scalars, we found that they are special cases of the conformally flat
plane-wave class (21.38) in ref. \cite{KSMH}, which can be written as
\begin{equation}
ds^{2} = 2d\zeta d\bar\zeta - 2dudv - 2H(\zeta,\bar\zeta,u)du^2, \label{cpw}
\end{equation}
where H is real,
\begin{equation}
H_{,\zeta\zeta} = H_{,\bar\zeta\bar\zeta} = 0,
\end{equation}
and where a bar and a comma denote complex conjugation and partial derivative,
respectively.

In the null tetrad
\begin{equation} \begin{array}{lcl}
\omega^{0} &=&
H(H_{,\zeta\bar\zeta})^{-1/2} du + (H_{,\zeta\bar\zeta})^{-1/2} dv, \\
\omega^{1} &=& (H_{,\zeta\bar\zeta})^{1/2} du, \\
\omega^{2} &=& d\bar\zeta, \\
\omega^{3} &=& d\zeta
\end{array} \label{cpwtetrada}
\end{equation}
the nonzero algebraically independent Cartan scalars ($C_n$) for this
spacetime may be written as
\begin{eqnarray}
\eb[12em][r]{$0^{\rm th}$ derivative:\hfill
$\tilde{\Phi}_{22'}$} & = & 1, \label{cpw0a} \\
 & & \nonumber \\
\eb[12em][r]{$1^{\rm st}$ derivative:\hfill
$\widetilde{\nabla\Phi}_{33'}$} & = &
-2\left((H_{,\zeta\bar\zeta})^{-1/2}\right)_{,u}, \label{cpw1a} \\
 & & \nonumber \\
\eb[12em][r]{$2^{\rm nd}$ derivative:\hfill
$\widetilde{\nabla^2\Phi}_{44'}$} & = &
(H_{,\zeta\bar\zeta})^{-2} H_{,\zeta\bar\zeta uu}, \label{cpw2a}
\end{eqnarray}
where we have used tilde for compatibility with the notation we have been
using in the type [(112)] analysis.

With a proper choice of the function $H$, the limit (\ref{l0a})--(\ref{l1a})
may be recovered. Indeed, equating (\ref{cpw1a}) to the constant value in
(\ref{l1a}), a particular solution for $H$ can be found, namely
\begin{equation}
H = 4\left(-2\frac{Q+1}{\sqrt{Q}}\right)^{-2}\zeta\bar\zeta u^{-2}. \label{H}
\end{equation}
All possible values of $\widetilde{\nabla\Phi}_{33'}$ in (\ref{l1a})
may be obtained from this solution. Therefore, the limit Cartan scalars
(\ref{l0a})--(\ref{l1a}) correspond to the line element (\ref{cpw}) with
$H$ given by eq. (\ref{H}). Note that the compatibility of the second
derivative (\ref{cpw2a}) does not need to be checked, since for
$\widetilde{\nabla\Phi}_{33'} = \mbox{const}$ the second derivative
is not an independent quantity.

\section{A static FRW solution as a third example} \label{rb}

We shall now study the limits of a static FRW model with closed spatial
section, which is a vacuum solution of Brans-Dicke equations with cosmological
constant, found by Romero and Barros
\cite{RomeroBarros1993a,RomeroBarros1993b}. Its line element and scalar field
may be written as
\begin{eqnarray}
ds^2 &=& dt^2 - A^2\left(dr^2 + sin^2r(d\theta^2 +
sin^2\theta d\varphi^2)\right), \\
\phi &=& \phi_0e^{\pm \sqrt{\frac{2\lambda}{2\omega+3}}\:t},
\end{eqnarray}
where
\begin{equation}
A  =  \sqrt{\frac{2\omega+3}{\lambda(\omega+1)}}\: , \label{rbA}
\end{equation}
\begin{equation}
\phi_0  \mbox{\ is a constant,}
\end{equation}
and $\lambda$ is the cosmological constant. In the null tetrad
\begin{equation} \begin{array}{lcl}
\omega^{0} &=& 1/\sqrt{2} (dt+Adr) , \\
\omega^{1} &=& 1/\sqrt{2} (dt-Adr), \\
\omega^{2} &=& 1/\sqrt{2}A\sin(r) (d\theta+i\sin(h) d\varphi) , \\
\omega^{3} &=& 1/\sqrt{2}A\sin(r) (d\theta-i\sin(h) d\varphi) ,
\end{array} \label{tetrada2}
\end{equation}
the nonzero algebraically independent Cartan scalars may be written as
\begin{eqnarray}
\eb[12em][r]{$0^{\rm th}$ derivative:\hfill
$\Phi_{00'}$} & = & 2\Phi_{11'}, \label{ee0a} \\
\eb[12em][r]{$\Phi_{11'}$} & = & \frac{1}{4}A^{-2}, \label{ee0b} \\
\eb[12em][r]{$\Phi_{22'}$} & = & 2\Phi_{11'}, \label{ee0c} \\
\eb[12em][r]{$\Lambda$} & = &  \Phi_{11'}. \label{ee0d}
\end{eqnarray}
This is a Petrov type 0 and Segr\'e type [1,(111)] metric. As there is no
function of the coordinates among the Cartan scalars, its a spacetime
homogeneous metric. Its isometry group is seven-dimensional and the isotropy
subgroup is the SO(3). The first order Cartan scalars vanish identically,
therefore the Karlhede algorithm stops. From (\ref{rbA}), as $\omega\ti$,
\begin{equation}
A\rA \sqrt{\frac{2}{\lambda}}. \label{Alim}
\end{equation}
The limit of the Cartan scalars are given by substituting $A$ in
(\ref{ee0a})--(\ref{ee0d}) for (\ref{Alim}). This is nothing but the
Einstein static model of the universe, which is a solution of the
general relativity field equations with cosmological constant (equals
to half of ours) and a matter distribution corresponding to dust.

In this case no other limit is possible since the Cartan scalars
do not depend on the coordinates.

\section{Conclusion} \label{con} \setcounter{equation}{0}

It was found that the Nariai metric may have three possible
limits as $\omega\ti$, namely: the Minkowski space, the FRW
metric with euclidian spatial section and the conformally flat
plane-wave solution (\ref{H}). Regarded as solutions of the
general relativity equations, these correspond to three
different matter distributions: vacuum, perfect fluid with a
equation of state $p=(\gamma'-1)\rho$ and a Segre type
[(112)] configuration.  The matter distribution corresponding to
the first and the third metrics differ from
that of Nariai's solution of Brans-Dicke theory.  On the other hand,
the second matter distribution is the same perfect fluid
as in Nariai as can be seen from eqs. (\ref{Qlim}) and
(\ref{Qlim2}).

The O'Hanlon-Tupper metric has the same limits as Nariai's.
Nevertheless, it is a vacuum solution of the Brans-Dicke field
equations. Therefore, the first limit, namely, the Minkowski
space, has the same matter configuration (vacuum), while this does not
occur with the other two limits.

The situation with the static vacuum solution of the Brans-Dicke
equations with cosmological constant discussed in
section~\ref{rb} is quite different. It has just one limit, namely,
the Einstein static metric, which regarded as a general relativity
solution cannot be obtained from the vacuum Einstein's equation, since
the Ricci spinor ($\Phi_{AB'}$) does not vanish.

In other words, it seems that, in  general, Brans-Dicke solutions do
not have a unique limit as $\omega\ti$. Thus, to say that a given
Brans-Dicke spacetime tends to a spacetime of general relativity may
sound meaningless.  Moreover, it may even happen, as the example in
section \ref{rb} explicitly shows, that none of the limits corresponds
to a general relativity solution with the same matter distribution.

\section*{Acknowledgments}

The authors thank the Conselho Nacional de Desenvolvimento
Cient\'{\i}fico e Tecnol\'ogico (CNPq) for financial support. F. M.
Paiva also thanks Marcelo Rebou\c{c}as, Antonio Teixeira and Graham
Hall for many useful discussions and the Departamento de F\'{\i}sica
da Universidade Federal da Para\'{\i}ba for the kind hospitality during
the final stage of this work.

\end{document}